\documentclass[aps,prd,preprint,preprintnumbers,unsortedaddress,superscriptaddress,showpacs,nofootinbib]{revtex4-1}
\usepackage{subfigure}
\usepackage{graphicx}
\usepackage{epsfig}
\usepackage{amsmath}
\usepackage{amsfonts}
\usepackage{amssymb}
\usepackage{xcolor}
\usepackage{slashed}
\usepackage{lipsum}
\usepackage{epstopdf}
\usepackage{hyperref}
\usepackage{float}
\usepackage{wrapfig}
\usepackage[utf8]{inputenc}
\usepackage{version}	
\hypersetup{colorlinks = true,
           allcolors = blue}

\newcommand{\tento}[1]{\times 10^{#1}}

\newcommand{\im}{{\rm Im}}

\newcommand{\br}{{\rm Br}}

\newcommand{\kev}{{\rm keV}}
\newcommand{\mev}{{\rm MeV}}
\newcommand{\gev}{{\rm GeV}}

\newcommand{\mC}{\mathcal{C}}

\newcommand{\mM}{\mathcal{M}}

\newcommand{\ddstar}{D^0\bar D^{*0}}

\newcommand{\itp}{\affiliation{CAS Key Laboratory of Theoretical Physics, Institute of Theoretical Physics, Chinese Academy of Sciences,  Zhong Guan Cun East Street 55, Beijing 100190, China}}
\newcommand{\ucas}{\affiliation{School of Physical Sciences, University of Chinese Academy of Sciences, Beijing 100049, China}}
\newcommand{\ific}{\affiliation{Departamento de F\'{\i}sica Te\'orica and IFIC, Centro Mixto Universidad de Valencia-CSIC Institutos de Investigaci\'on de Paterna, Aptdo. 22085, 46071 Valencia, Spain}}
\allowdisplaybreaks[1]

\newcommand{\mN}{\mathcal{N}}
\newcommand{\cc}{{\rm c.c.}}

\begin{document}
\title{Triangle singularity in the \texorpdfstring{$B^-\to K^-\pi^0X(3872)$}{BmtoKmpi0X} reaction and sensitivity to the \texorpdfstring{$X(3872)$}{X} mass}
\date{\today}
\author{Shuntaro~Sakai}
\email{shsakai@mail.itp.ac.cn}
\itp
\author{Eulogio~Oset}
\email{oset@ific.uv.es}
\ific
\author{Feng-Kun~Guo}
\email{fkguo@itp.ac.cn}
\itp
\ucas

\begin{abstract} 
We have done a study of the $B^- \to K^- \pi^0 X(3872)$ reaction by means of a triangle mechanism via the chain of reactions: $B^- \to K^- D^{*0}\bar D^{*0}$; $D^{*0} \to \pi^0 D^0$; $D^0\bar D^{*0}\to X(3872)$.
We show that this mechanism generates a triangle singularity in the $\pi^0 X(3872)$ invariant mass for a very narrow window of the $X(3872)$ mass, 
around the present measured values, 
and show that the peak positions and the shape of the mass distributions are sensitive to the $X(3872)$ mass, 
such that a measurement of the reaction can serve to improve on the present values of this mass. In particular, we point out that the $X(3872)$ mass relative to the $D^0\bar D^{*0}$ threshold may be extracted from the asymmetry of the $\pi^0X$ line shape.
\end{abstract}
\maketitle

\section{Introduction}

Triangle singularities (TSs) in physical processes, 
introduced in the 1960s~\cite{karplus,landau} and searched for in some reactions without much success~\cite{booth,anisovich},
have undergone a spectacular revival in recent years.
The TS in the physical region
stems from mechanisms that involve three intermediate particles, which can be placed simultaneously on shell while being collinear,
and represent a process that can occur classically (Coleman-Norton theorem~\cite{coleman}).
With the outcome of a wealth of experimental data at present, examples of enhancements in cross sections due to TSs have become available.
One of the early examples was the explanation of the peak observed by the COMPASS Collaboration in the $\pi f_0(980)$ final state at 1420~MeV, attributed in Ref.~\cite{compass} to a new resonance $a_1(1420)$.
It was suggested in Refs.~\cite{qiangz,mikha,acetidai} that the peak corresponded to the $\pi f_0(980)$ decay mode of the $a_1(1260)$ due to a TS.
Another example was the explanation due to a TS \cite{wu,acetiliang,wuwu,acha,shesta} of the spectacular enhancement of the isospin forbidden $\eta(1405)$ decay into $\pi^0f_0(980)$ compared to $\pi^0a_0(980)$~\cite{beseta}.

Other recent examples of TSs affecting reactions include the explanation of the $\pi N(1535)$ production \cite{vinisakai} in the $\gamma p\to p\pi^0\eta$ reaction \cite{gutz},
the explanation of the enhancement of the cross section of the $\gamma p\to K^+\Lambda(1405)$ reaction around $\sqrt{s}=2110$~MeV~\cite{moriya} given in Ref.~\cite{wangguo}, and the clarification of the supposed $\pi a_0(980)$ decay mode of the $f_1(1420)$ resonance, which is suggested in Ref.~\cite{viniaceti} to come from the $f_1(1285)$ decay into this mode via a TS.
The main decay mode, $K^*\bar K$, of the ``$f_1(1420)$'' comes from the normal $f_1(1285)$ decay mode in this channel, which appears at a higher invariant mass than the nominal mass, 1285~MeV, when the $K^*$ is placed on shell.
For an extensive review on various manifestations of TSs in hadronic reactions, we refer to Ref.~\cite{guosakai}.

A new reformulation of the problem of TSs is done in Ref.~\cite{bayarguo} with a very simple formula to determine whether a TS is in the physical 
region
and where it appears, and an extended review on this issue can be seen in Ref.~\cite{guosakai}.
It is shown there that some reactions involving TSs, which are in the vicinity of the physical region, serve to enhance the production of hadronic molecules.
This would be the case of the $X(4260)\to \gamma X(3872)$ proposed in Ref.~\cite{guohanhart} and posteriorly measured at BESIII~\cite{guobes}.
Another case is given by the $B^-\to D^{*0}\pi^-f_0(980)$ $[a_0(980)]$, which for some energy of the $\pi^-f_0$ $(a_0)$ system develops a singularity \cite{pavaosakai}.
In some cases, it also serves to enhance the isospin-forbidden production modes like in the $D_s^+\to\pi^+\pi^0a_0(980)$ $[f_0(980)]$ decay, where the isospin-forbidden $f_0(980)$ production mode is enhanced compared to the isospin-allowed $a_0(980)$ mode due to a TS~\cite{Sakai:2017iqs}.
A detailed list of the proposed reactions has been tabulated in Table~2 of Ref.~\cite{guosakai}.

Concerning the $X(3872)$ production, an interesting proposal was made to observe this resonance and measure its mass with high precision~\cite{fengkun}.
The reaction is $D^{*0}\bar D^{*0}\to \gamma X(3872)$, where the $D^{*0}\bar D^{*0}$ would be produced by some short-distance source.
The reaction develops a TS at $\gamma X$ invariant masses which are very sensitive to the $X(3872)$ mass relative to the $\ddstar$ threshold, i.e., the $X(3872)$ binding energy,
\begin{align}
\delta_X = m_{D^0} + m_{D^{*0}} -m_X,\label{eq:be}
\end{align}
and could serve to greatly improve on the present uncertainties in the $X(3872)$ binding energy.
When the $D^{*0}\bar D^{*0}$ are in $S$ wave, the $\gamma X$ line shape is characterized by a $D^{*0}\bar D^{*0}$ threshold cusp and a TS-induced peak above the cusp. As a result, the line shape is extremely sensitive to the $X(3872)$ binding energy. 
A possible reaction implementing this mechanism with the $D^{*0}\bar D^{*0}$ being in $P$ wave has been given in Refs.~\cite{braaten,Braaten:2019gwc} with the $e^+e^-\to \gamma X(3872)$ process.
A variant of this reaction is also proposed in Ref.~\cite{brapion} with the $B^0\to K^+ X\pi^-$ and $B^+\to K^0 X\pi^+$ processes, where the $X\pi$ system develops a TS.
The mechanism for the production is given by $B\to KD^{*}\bar D^{*0}$ $(B\to K\bar D^{*}D^{*0})$ followed by $D^*\to\pi D^0$ $(\bar D^*\to\pi \bar D^0)$ and $\bar D^{*0}D^0$ $(D^{*0}\bar D^0)$ fusing into the $X(3872)$.

The present work deals again on the $B\to KX\pi$ reaction with a different emphasis, which is to show the sensitivity of the $X\pi$ invariant mass distribution to the TS position and to the $X(3872)$ mass.
Since we aim at high precision in the determination of the TS energy, we must improve on approximations done in Ref.~\cite{brapion}.
The methods used and input are also different, and we also discuss the need to convolve the results with the mass distribution of the $X(3872)$ to compare with experiment.
In addition, we focus on the sensitivity of the $\pi^0X$ line shape on the $X(3872)$ binding energy.

This paper is organized as follows.
The formalism and some general discussions on the TS are presented in 
Sec.~\ref{sec:for}. The amplitude for the triangle diagrams is evaluated in 
Sec.~\ref{sec:eva}. In 
Sec.~\ref{sec:res}, we present the numerical results and the related discussions. In particular, we suggest that the $X(3872)$ binding energy may be extracted from the asymmetry of the $\pi^0X$ line shape around the TS. A brief summary is given in 
Sec.~\ref{sec:conc}.

\section{Formalism}
\label{sec:for}
\subsection{General considerations}

We will study the $B^-\to K^-\pi^0X(3872)$ reaction which proceeds via the diagrams depicted in Fig.~\ref{fig1}.
\begin{figure}[tb]
 \centering
 \includegraphics[width=15cm]{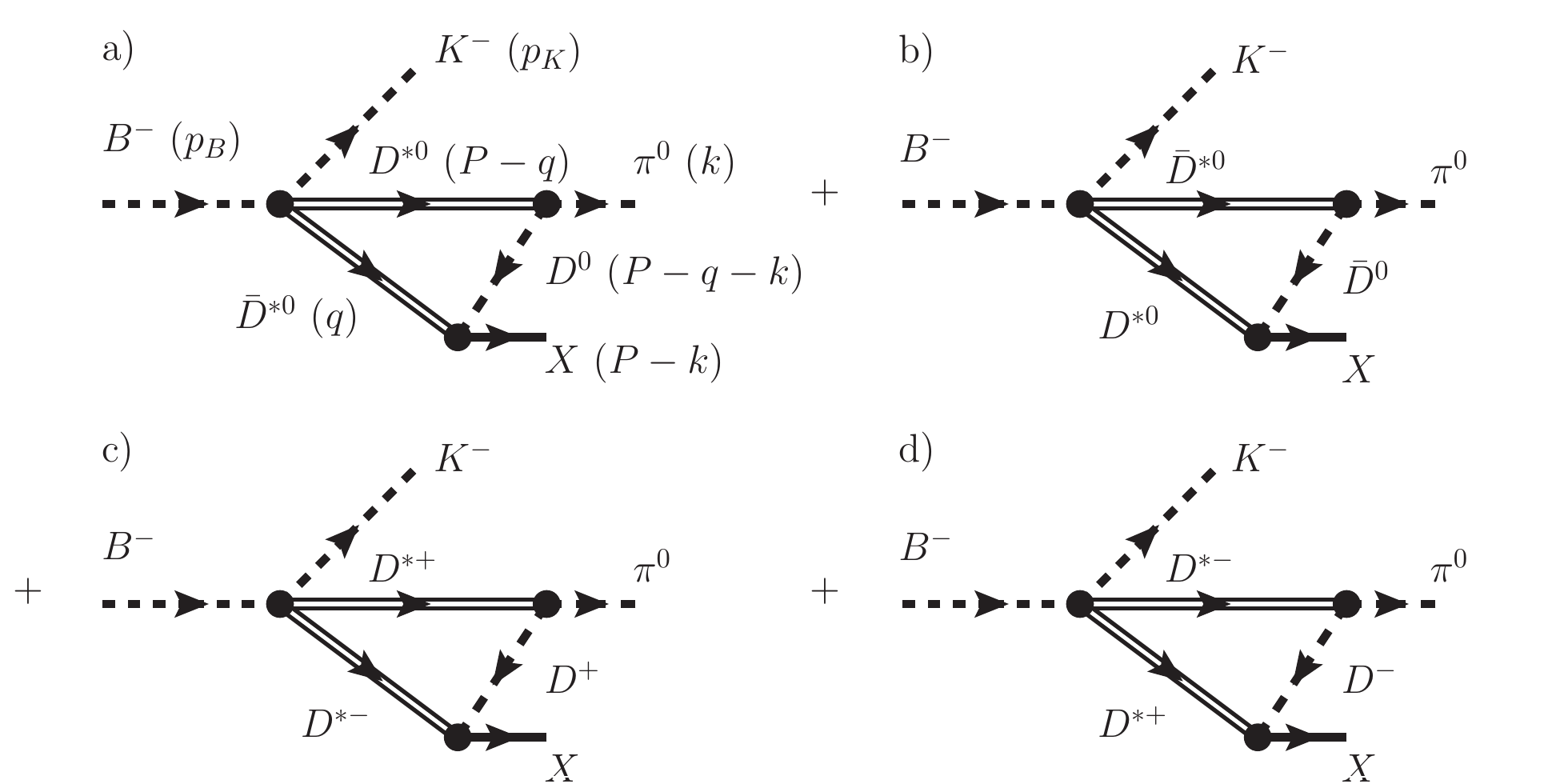}
 \caption{Triangle diagrams leading to $K^-\pi^0X$ in $B^-$ decay.
 The momenta of the corresponding particles are in parentheses.
}
 \label{fig1}
\end{figure}

The picture implicitly uses the fact that the $X(3872)$ couples strongly to $D^*\bar D+\cc$. This is an observed fact as the $D^0\bar D^{*0}+\cc$ mode consists of a large portion of the $X(3872)$ decays~\cite{pdg,changzheng,Braaten:2019ags}. Although one does not need to assume the $X(3872)$ to be a $D\bar D^*+\cc$ hadronic molecule, a strong coupling is natural under such an assumption.
In pictures as in Refs.~\cite{gamermann,juan} where the $D\bar D^*$ interaction is studied, the $X(3872)$ results in a dynamically generated resonance from the interaction and the coupling of $X(3872)$ to the $D\bar D^*+\cc$ components can be computed~\cite{gamermann,isodani,juan}.
In the XEFT effective theory, it is treated as a $D^0\bar D^{*0}$ molecule~\cite{mehen}. For recent reviews of the hadronic molecular description of the $X(3872)$, see \cite{Guo:2017jvc,Kalashnikova:2018vkv,Brambilla:2019esw}.

In Ref.~\cite{bayarguo}, it is discussed that the TS appears when the $D^*,\bar D^*$ are placed on shell in the loops of Fig.~\ref{fig1} simultaneously with the $D,\bar D^*$ $(\bar D,D^*)$.
In addition, in the diagram of Fig.~\ref{fig1}a), the $D^{*0}$ and $\pi^0$ have to have their momentum in the same direction in the $\pi^0X$ 
center-of-mass (c.m.)
frame.
These conditions lead to two solutions for the momenta of the $\bar D^{*0}$ in Fig.~\ref{fig1}a), and it is shown that only one (satisfying the Coleman-Norton theorem~\cite{coleman}) leads to a TS in the physical region.
These conditions are summarized in a very easy equation,
\begin{align}
 q_{a-}=q_{\rm on},\label{eq:5-1}
\end{align}
where $q_{\rm on}$ and $q_{a-}$ are the momenta of the intermediate particle that connect the initial and final heavy particles evaluated by putting the $D^*\bar D^*$ and $D^*\bar D$ (or $\bar D^* D$) on shell, respectively. The expressions for $q_{a-}$ and $q_{\rm on}$ are given in Ref.~\cite{bayarguo}.
Let us take a mass for the $X$, $M_X=3871.70~\mev$, inside the currently measured range, $3871.69\pm0.17$~MeV~\cite{pdg}, and apply the condition of Eq.~\eqref{eq:5-1}.
We obtain the TS at an $\pi^0X$ invariant mass $M_{\pi^0X}=4013.84~\mev$.
It is interesting to see that the there is a very narrow window of masses where the TS appears. The range is given by [see Eq.~(59) in Ref.~\cite{guosakai}]
\begin{equation}
    M_X\in \left[m_{D^0}+m_{D^{*0}},\, \sqrt{2 \left(m_{D^{*0}}^2+m_{D^0}^2 \right) -m_{\pi^0}^2 } \right] = [3871.68,\, 3871.93]~\mev,
    \label{eq:mx}
\end{equation}
and, with Eq.~(60) in Ref.~\cite{guosakai}, correspondingly, the TS is located in the range,
\begin{equation}
    M_{\pi^0X} \in \left[ 2 m_{D^{*0}},\, \sqrt{ m_{D^{*0}} \left( 2 m_{D^{*0}} + m_{D^0} + \frac{m_{D^{*0}}^2-m_{\pi^0}^2}{m_{D^0}} \right) } \right] = [4013.70,\, 4013.96]~\mev, \label{eq:mpix}
\end{equation}
where the central Particle Data Group (PDG) FIT values of the involved meson masses, $m_{D^0}=1864.83$~MeV and $m_{D^{*0}}=2006.85$~MeV~\cite{pdg}, are used.

The above ranges can be compared with those for the reaction proposed in Ref.~\cite{fengkun}. They can be obtained by simply replacing $m_{\pi^0}$ by the zero photon mass,
\begin{equation}
    M_X \in [3871.68,\, 3874.28]~\mev,\quad M_{\gamma X}\in [4013.70,\, 4016.40]~\mev.
\end{equation}
The ranges in Eqs.~\eqref{eq:mx} and \eqref{eq:mpix} are much narrower, but fortunately, the narrow window in Eq.~\eqref{eq:mx} precisely covers the region where the mass of the $X(3872)$ lies at present, with $M_X=3871.69\pm 0.17~\mev$~\cite{pdg}.
Because the 
$\pi^0 X$ 
line shape depends strongly on the $X(3872)$ binding energy, this gives hope that, investigating precisely where the singularity peak appears in the $B\to K\pi X$ decays, one can still induce in retrospective the mass of the $X(3872)$ with much precision.
The fact that Eq.~\eqref{eq:5-1} has no solution outside the narrow window discussed above does not mean that there is no enhancement in the mass distribution of the $\pi X$.
By inertia, the enhancement remains, even if the intermediate particles in the loops are slightly off shell.
But the intensity of the peak and its shape differs from the one in the narrow window of Eq.~\eqref{eq:5-1} because in such a case, the TS is located in the complex $M_{\pi^0X}$ plane even if the $D^{*0}$ width is neglected.

We aim at an extraction of the $X(3872)$ binding energy with a high precision for which a very precise calculation for the shape of the $\pi^0X$ invariant mass distribution must be done.
This means more precision than the one assumed in Ref.~\cite{brapion}, where several kinematical simplifications are done, among them treating the pion nonrelativistically. In order to get a simple analytic expression for the amplitude, the approximation in Ref.~\cite{brapion} amounts to the approximation of the scalar triangle loop integral by an expression,
\begin{equation}
    \propto \frac1{\sqrt{q^2/(2m_{\pi^0}) -\delta_0 - \delta_X + i\Gamma_{D^{*0}} } + \sqrt{-\delta_X + i\Gamma_{D^{*0}}/2} } ,
\end{equation}
where $q$ is the size of the pion momentum in the $\pi X$ 
c.m.
frame, and $\delta_0 = m_{D^{*0}} - m_{D^0} - m_{\pi^0}$.
However, such an approximation changes the singularity behavior. The TS is logarithmic, while the above one is certainly not. Furthermore, the square-root branch point at the $M_{\pi^0X}=2m_{D^{*0}}$ when 
$\Gamma_{D^{*0}}=0$
is also lost. Therefore, the approximation in Ref.~\cite{brapion} may be used for an estimate of the reaction rate, but is not applicable for a precise description of $\pi^0X$ line shape in the vicinity of the TS.
In fact, the same approximation has been used to establish the power counting in a nonrelativistic effective field theory~\cite{Guo:2012tg}, and it has been pointed out in Ref.~\cite{Guo:2017jvc} (see Sec.~IV.A.2 therein) that the expansion can only be made in a region away from the TS (for a discussion of the TS in the nonrelativistic formalism, we refer to Ref.~\cite{Guo:2014qra}).

One may wonder what happens with the diagrams of Figs.~\ref{fig1}c) and \ref{fig1}d).
One can see using Eq.~\eqref{eq:5-1} that they do not develop a singularity in the physical region using the actual mass of the $X(3872)$.
Instead, we have to go to an $X(3872)$ mass around 3880.00~MeV for the TS to appear at $M_{\pi^0X}=4020.54$~MeV.
It is clear that this singularity will not have any relevance in the region of the actual $X(3872)$ masses, and the process will be driven by the contribution of diagrams a) and b) of Fig.~\ref{fig1}.
This means that even if in the picture of Refs.~\cite{gamermann,isodani}, the $X(3872)$ appears from the coupled channels, $D^{*0}\bar D^0+\cc$ and $D^{*+}D^-+\cc$, in this reaction, the charged channels do not play any role in the TS.
In contrast, there are other reactions where the charged components are essential to understand the experimental results, as in the $X(3872)\to J/\psi \gamma$ decay, as shown in Refs.~\cite{fraraquel,Guo:2014taa}.

The dominance of diagrams a) and b) of Fig.~\ref{fig1} in the TS has become clear from the former discussion, 
but there are extra reasons that make the charged mechanisms further irrelevant, and for this, we go to the root of the weak decay that produces $B^-\to K^-D^{*0}\bar D^{*0}$ and $B^-\to K^-D^{*+}D^{*-}$. 
The $B^-\to K^-D^{*0}\bar D^{*0}$ can proceed via external emission \cite{chau} and hadronization as shown in Fig.~\ref{fig2}.
\begin{figure}[t]
 \centering
 \includegraphics[width=12cm]{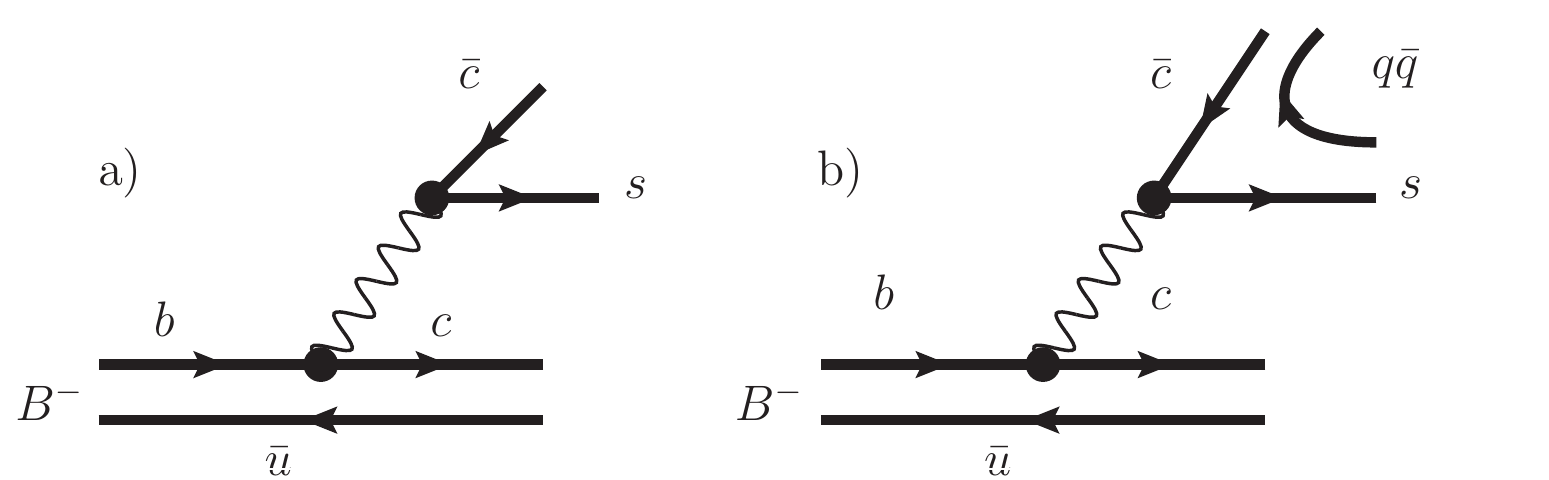}
 \caption{a)~External emission of $B^-$ decay at the quark level.
 b)~Hadronization of the $s\bar{c}$ pair.}
 \label{fig2}
\end{figure} 
Considering $\bar{q}q\equiv \bar{u}u+\bar{d}d+\bar{s}s$, the hadronization of diagram \ref{fig2}b) leads to
\begin{align}
 K^-\bar D^{*0}+\bar K^0D^{*-}
\end{align}
plus other components that we are not interested in.
The $c\bar{u}$ is the $D^{*0}$, and then we can have the $K^-D^{*0}\bar D^{*0}$ component but not the $K^-D^{*+}D^{*-}$.
To create the $K^-D^{*+}D^{*-}$ component, we must resort to the internal emission~\cite{chau} as shown in Fig.~\ref{fig3}.
\begin{figure}[t]
 \centering
 \includegraphics[width=12cm]{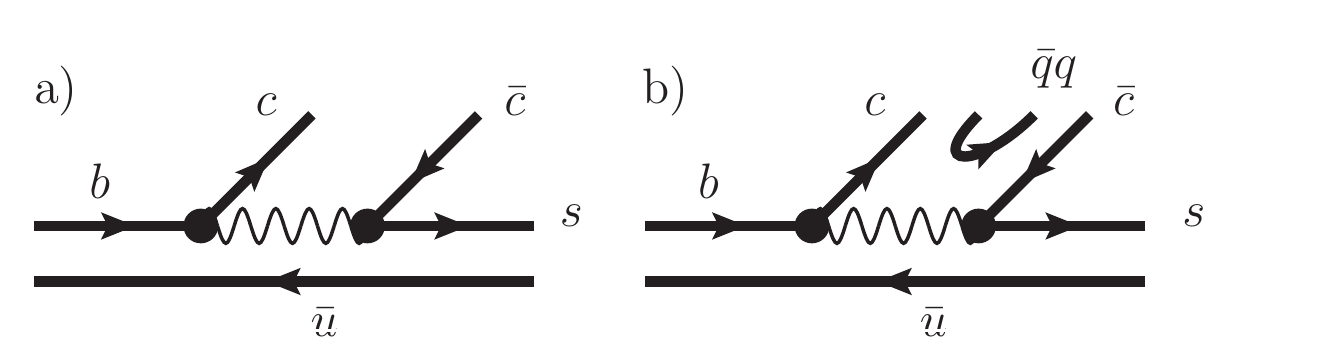}
 \caption{a)~Internal emission of $B^-$ decay at the quark level.
 b)~Hadronization of the $c\bar{c}$ pair.}
 \label{fig3}
\end{figure} 
Now, the $c\bar{c}$ component through $\sum_ic\bar{q}_iq_i\bar{c}$ in the hadronization process leads to 
\begin{align}
 K^-(D^{*0}\bar D^{*0}+D^{*+}D^{*-})
\end{align}
among other components.
The $K^-D^{*+}D^{*-}$ decay is possible through the internal emission, which is dynamically suppressed by a color factor \cite{chau}.
Then, it is not surprising to see the experimental results~\cite{pdg},
\begin{align}
 \br(B^-\to K^-D^{*0}\bar D^{*0})=(1.12\pm 0.13)\tento{-2},\label{eq:8-1}\\
 \br(B^-\to K^-D^{*+}D^{*-})=(1.32\pm 0.18)\tento{-3}.\label{eq:8-2}
\end{align}
We shall use the datum of Eq.~\eqref{eq:8-1} to estimate the vertex $B^-\to K^-D^{*0}\bar D^{*0}$ needed for the evaluation of the diagrams of Fig.~\ref{fig1}.

\subsection{The \texorpdfstring{$B^-\to K^-D^{*0}\bar D^{*0}$}{BmtoKmDst0Dst0bar} reaction}

There is work done on the $B\to KD^{(*)}\bar D^{(*)}$ reactions, parametrizing the amplitudes and making a fit to experimental data,
relating different decays~\cite{Zito:2004kz,poireau}.
Since we are only concerned about the $B^-\to K^-D^{*0}\bar D^{*0}$ decay, we can make the analysis for this channel alone using the information of Eq.~\eqref{eq:8-1}.
For an order-of-magnitude estimate of the reaction rate, we neglect the contribution from possible intermediate resonances, which is irrelevant for the TS. Since we are interested in the $\pi^0X$ line shape in the immediate vicinity of the $D^{*0}\bar D^{*0}$ threshold, we choose the amplitude involving the lowest angular momentum between $D^{*0}$ and $\bar D^{*0}$,
\begin{align}
 t_{B^-\to K^-D^{*0}\bar D^{*0}}=\mC\,\vec{\epsilon}_{D^{*0}}\cdot\vec{\epsilon}_{\bar D^{*0}},
\end{align}
{where $\mC$ is a constant to be determined from experimental data, and $\vec{\epsilon}_{D^{*0}}$ and $\vec{\epsilon}_{\bar D^{*0}}$ are the $D^{*0}$ and $\bar D^{*0}$ polarization vectors, respectively.}
It is unnecessary to take the $\epsilon^0$ component in such a case as shown in Ref.~\cite{sakairamos} (see the Appendix of this reference).
We have in this case,
\begin{align}
 \frac{d\Gamma_{B^-\to K^-D^{*0}\bar D^{*0}}}{dM_{D^{*0}\bar D^{*0}}}=&\frac{p_{K^-}\tilde{p}_{D^{*0}}}{(2\pi)^34M_{B^-}^2}\overline{|t_{B^-\to K^-D^{*0}\bar D^{*0}}|^2}\notag\\
 =&\frac{p_{K^-}\tilde{p}_{D^{*0}}}{(2\pi)^34M_{B^-}^2}3\mC^2,\label{eq:dgambkdstdstb}
\end{align}
with
{$p_{K^-}$, the magnitude of the three momentum of $K^-$ in the $B^-$ rest frame, and $\tilde{p}_{D^{*0}}$, that of $D^{*0}$ in the $D^{*0}\bar D^{*0}$ c.m. frame, given by}
\begin{align}
 p_{K^-}=&\frac{1}{2M_{B^-}}\lambda^{1/2}(M_{B^-}^2,m_{K^-}^2,M^2_{D^{*0}\bar D^{*0}}),\\
 \tilde{p}_{D^{*0}}=&\frac{1}{2M_{D^{*0}\bar D^{*0}}}\lambda^{1/2}(M^2_{D^{*0}\bar D^{*0}},m_{D^{*0}}^2,m_{\bar D^{*0}}^2),
\end{align}
where $\lambda(x,y,z)=x^2+y^2+z^2-2xy-2yz-2zx$. 
Then we determine $\mC^2$ as 
\begin{align}
 \frac{\mC^2}{\Gamma_{B^-}}=\br(B^-\to K^-D^{*0}\bar D^{*0})\frac{1}{\int\frac{dM_{D^{*0}\bar D^{*0}}}{(2\pi)^3}p_{K^-}\tilde{p}_{D^{*0}}\frac{3}{4M_{B^-}^2}},
\end{align}
{which follows from Eq.~\eqref{eq:dgambkdstdstb}, using the experimental data for the branching ratio from Eq.~\eqref{eq:8-1}.} 
Let us emphasize that this value is only for an estimate of the reaction rate and does not play any role in the $\pi^0X$ line shape.

\subsection{The \texorpdfstring{$D^{*0}\to D^0\pi^0$}{Dst0toD0pi0} amplitude}
{Denoting the coupling constant of $D^{*0}\to \pi^0D^0$ by $\tilde{g}$,}
we write 
\begin{align}
 -it_{D^{*0}\to \pi^0D^0}=-\frac{i\tilde{g}}{\sqrt{2}}(\vec{p}_{\pi^0}-\vec{p}_{D^{*0}})\cdot\vec{\epsilon}_{D^{*0}}.\label{eq:10-2}
\end{align}
{We employ isospin symmetry to relate the partial width of $D^{*0}\to \pi^0D^0$ with that of $D^{*+}\to \pi^+ D^0$.}
Using the $D^{*+}$ full width, the $D^{*+}\to \pi^+ D^0$ branching fraction, 
and the fact that the branching ratio of $D^{*0}\to \pi^0D^0$ is $64.7\%$, we obtain~\cite{fengkun}
\begin{align}
 \tilde{g}=8.40;~\Gamma_{D^{*0}}=55.4~\kev.
\end{align}
The value of 
$55.4$~keV
is similar to the one obtained in Ref.~\cite{Rosner:2013sha} and the one used in Ref.~\cite{bradstar}, $\Gamma_{D^{*0}}\sim 60~\kev$.
The amplitude for $\bar{D}^{*0}\to \pi^0\bar{D}^0$ is obtained changing the direction of the lines in the $D^{*0}$ decay, and hence, one obtains the same coupling as in Eq.~\eqref{eq:10-2} with an opposite sign.

\subsection{The \texorpdfstring{$X(3872)\to D^{*0}\bar D^{0}$}{XtoDst0Dst0} coupling}

Here, we make connection with Refs.~\cite{gamermann,isodani} where, referring to couplings, or wave functions at the origin, the $X(3872)$ is written as
\begin{align}
 X=\frac{1}{2}(D^{*+}D^-+D^{*0}\bar D^0-D^{*-}D^+-\bar D^{*0}D^0),\label{eq:10-1}
\end{align}
with our phase convention for the doublets $(D^+,-D^0)$, $(\bar D^0,D^-)$ (same for $D^*$), and  $CD^+=D^-$, $CD^{*+}=-D^{*-}$ for the $C$-parity operator.
The wave function of Eq.~\eqref{eq:10-1} has $I^G(J^{PC})=0^+(1^{++})$, as by the PDG~\cite{pdg}.
In Ref.~\cite{isodani}, the coupled channels are studied considering mass differences between the neutral and charged channels, 
and the couplings to the individual channels are very close to what we obtain from Eq.~\eqref{eq:10-1}. 
The coupling of the $X(3872)$ to $D^{*0}\bar D^0$ is given by
\begin{align}
 -it_{X,D^{*0}\bar D^0}=-\frac{ig_{X}}{2}\vec{\epsilon}_{D^{*0}}\cdot\vec{\epsilon}\,_X,\label{eq:11-1}
\end{align}
where by $g_X$, we mean the coupling of the $X(3872)$ to the whole combination of Eq.~\eqref{eq:10-1}.\footnote{Note that the coupling to $\bar D^{*0}D^0$ is just opposite to Eq.~\eqref{eq:11-1}. {We should mention that in the literature, there are different versions for the  $D^* \bar D+\cc$ molecule that imply different couplings of the $X(3872)$ to the charged and neutral components~\cite{Swanson:2003tb,Braaten:2006sy,Dong:2009uf,isodani,Dong:2009yp,Guo:2014hqa}. 
A thorough discussion on the meaning of the couplings, wave functions at the origin, and probabilities of the charged and neutral components is made in Ref.~\cite{isodani}, where the couplings are found to be very close to those implied by the combination of Eq.~\eqref{eq:10-1}.  While these different options are relevant in the study of some processes, in the present case, we have shown that there is only relevant contribution from the neutral components, for which Eq.~\eqref{eq:11-2} is a reasonable approximation. 
Once again, some changes in this coupling affect the strength of the widths obtained but not the position of the peaks and shapes, which is the relevant output of our calculations.
}}
One can make an estimate of $g_X$ by using the Weinberg compositeness condition \cite{weinberg,baru,danijuan} with the normalization required here~\cite{danijuan},\footnote{A similar relationship of the coupling and binding energy is used in Ref.~\cite{Lin:2017mtz} to estimate the coupling of $P_c$ to $\bar D^{(*)}\Sigma_c^{(*)}$.}
\begin{align}
 g_X^2=&\frac{16\pi s}{\mu}\sqrt{2\mu \delta_X},\label{eq:11-2}
\end{align}
where $s$ is the square of the mass of the $X(3872)$, $\mu$ the reduced mass of $D^{*0}$ and $D^0$, and $\delta_X$ the binding energy of the $X(3872)$ with respect to the $D^{*0}\bar D^0$ threshold defined in Eq.~\eqref{eq:be}.
For different binding energies, we find
\begin{align}
 \begin{split}
  g_X=&2.77~\gev~{\rm with}~\delta_X=50~\kev,\\
  g_X=&3.29~\gev~{\rm with}~\delta_X=100~\kev,\\ 
  g_X=&3.64~\gev~{\rm with}~\delta_X=150~\kev.
 \end{split}
 \label{eq:11-3}
\end{align}
We should note that Eq.~\eqref{eq:11-2} is valid for bound states.
We shall also consider the cases where the $D^{*0}\bar D^0$ pair is not bound. It can be computed as the residue of the scattering $T$ matrix in the same way as the bound state case with the $X(3872)$ mass used as an input.
In this case, we should recall that the $D^{*+}D^-$ and $D^{*-}D^+$ components are bound by about 8~MeV, and when one works with coupled channels,
as in Refs.~\cite{gamermann,isodani}, there is basically no difference whether the $D^{*0}\bar D^0$ channel is bound or unbound, and the coupling can be equally calculated in either case.
All this said, for the evaluations, we shall work with  a coupling,
\begin{align}
 g_X=3~\gev,
\end{align}
close to the values found in Eq.~\eqref{eq:11-3} and in Ref.~\cite{isodani}.
Once again we should note that the approximations that we have done only affect the strength, but not the position of the TS or the $\pi^0X$ line shape.

We have now all the ingredients needed for the evaluation of the triangle diagrams of Fig.~\ref{fig1}, which we do in the next section.

\section{Evaluation of the triangle diagrams}
\label{sec:eva}

We are now in a condition to evaluate the diagrams of Figs.~\ref{fig1}a) and \ref{fig1}b).
First, we should note that the two diagrams give exactly the same contribution,
since, in Fig.~\ref{fig1}b), there are two minus signs relative to Fig.~\ref{fig1}a) from the $D^{*0}\to \pi^0D^0$ $(\bar D^{*0}\to\pi^0\bar D^0)$
and $\bar D^{*0}D^0\to X$ $(D^{*0}\bar D^0\to X)$ vertices.
Hence, we just have to evaluate the diagram of Fig.~\ref{fig1}a).
The amplitude (including the factor 2) is given by
\begin{align}
 -i\mM_{B^-\to K^-\pi^0X}=&\; 2\int\frac{d^4q}{(2\pi)^4}\left[+\frac{ig_{X}}{2}\vec{\epsilon}_X\cdot\vec{\epsilon}_{\bar D^{*0}}\right]\left[-\frac{i\tilde{g}}{\sqrt{2}}(\vec{p}_{\pi^0}-\vec{p}_{D^0})\cdot\vec{\epsilon}_{D^{*0}}\right]\left[-i\mC\vec{\epsilon}_{D^{*0}}\cdot\vec{\epsilon}_{\bar D^{*0}}\right]\notag\\
 &\frac{i}{(P-q)^2-m_{D^{*0}}^2+i\epsilon}\frac{i}{q^2-m_{\bar D^{*0}}^2+i\epsilon}\frac{i}{(P-q-k)^2-m_{D^{0}}^2+i\epsilon} \notag\\
 =&\; i\frac{g_X\tilde{g}\mC}{\sqrt{2}}({\epsilon}_X)_i \left(i\int\frac{d^4q}{(2\pi)^4}(2{k}+{q}\,)_i\right.\notag\\
 &\left.\frac{1}{(P-q)^2-m_{D^{*0}}^2+i\epsilon}\frac{1}{q^2-m_{\bar D^{*0}}^2+i\epsilon}\frac{1}{(P-q-k)^2-m_{D^{0}}^2+i\epsilon}\right)\notag\\
 \equiv&\; i\frac{g_X\tilde{g}\mC}{\sqrt{2}}({\epsilon}_X)_i (t_T)_i,\label{eq:13-1}
\end{align}
where the evaluation has been done in the $\pi^0X(3872)$ 
c.m.
frame, where $\vec{P}=\vec{0}$.

In the next step, we evaluate $(t_T)_i$ of Eq.~\eqref{eq:13-1}, by performing analytically the $q^0$ integration.
Given the fact that in the TS all intermediate particles of the loops are placed on shell, it is sufficient to take just the positive energy part of the meson propagators, and we find
\begin{align}
 (t_T)_i=&\,i\int\frac{d^4q}{(2\pi)^4}\frac{1}{(P-q)^2-m_{D^{*0}}^2+i\epsilon}\frac{1}{q^2-m_{\bar D^{*0}}^2+i\epsilon}\frac{1}{(P-q-k)^2-m_{D^{0}}^2+i\epsilon}(2{k}+{q})_i\notag\\
 \simeq&\,i\int\frac{d^4q}{(2\pi)^4}\frac{(2{k}+{q})_i}{8E_{D^{*0}}E_{\bar D^{*0}}E_{D^{0}}}\frac{1}{P^0-q^0-E_{D^{*0}}+i\epsilon}\frac{1}{q^0-E_{\bar D^{*0}}+i\epsilon}\frac{1}{P^0-q^0-k^0-E_{D^{0}}+i\epsilon}\notag\\
 =&\int\frac{2\pi q^2dqd\cos\theta}{(2\pi)^3}\frac{(2{k}+{q})_i}{8E_{D^{*0}}E_{\bar D^{*0}}E_{D^{0}}}\frac{1}{W-E_{\bar D^{*0}}-E_{D^{*0}}+i\epsilon}\frac{1}{W-E_{\bar D^{*0}}-k^0-E_{D^{0}}+i\epsilon},\label{eq:tTint}
\end{align}
where the energies of $D^{*0}$, $\bar D^{*0}$, and $D^0$ in the $\pi^0X(3872)$ c.m. frame are given by $E_{D^{*0}}=\sqrt{q^2+m_{D^{*0}}^2}$, $E_{\bar D^{*0}}=\sqrt{q^2+m_{\bar D^{*0}}^2}$, and $E_{D^{0}}=\sqrt{(\vec{q}+\vec{k})^2+m_{D^{0}}^2}$, respectively, and $\theta$ is the angle between $\vec{q}$ and $\vec{k}$: $\vec{q}\cdot\vec{k}=qk\cos\theta$.
The $q^0$ integral was done using Cauchy's theorem by picking up the pole at $q^0=E_{\bar D^{*0}}-i\epsilon$ in the lower half of the complex $q^0$ plane.
Here, $W$ denotes $P^0=M_{\pi^0X}$.
The magnitudes of the three momentum ($k$) and energy ($k^0$) of the external $\pi^0$ are given by
\begin{align}
 k=&\frac{1}{2M_{\pi^0X}}\lambda^{1/2}(M_{\pi^0X}^2,m_{\pi^0}^2,M_X^2),~
 k^0=\sqrt{k^2+m_{\pi^0}^2}=\frac{1}{2M_{\pi^0X}}(M_{\pi^0X}^2+m_{\pi^0}^2-M_X^2). 
\end{align}
We take into account the width of $D^{*0}$ by the following replacement of the $D^{*0}$ and $\bar D^{*0}$ energies in the integrand of Eq.~\eqref{eq:tTint}:
\begin{align}
 E_{D^{*0}}\to& E_{D^{*0}}-i\Gamma_{D^{*0}}/2,\\
 E_{\bar D^{*0}}\to& E_{\bar D^{*0}}-i\Gamma_{\bar D^{*0}}/2. 
\end{align}

Because $\vec{k}$ is the only vector quantity which is not integrated, the vector integral $\int \vec{q}D_\Delta$ should be proportional to $\vec{k}$~\cite{bayarguo,Aceti:2015zva},
\begin{align}
 &\int \frac{d^3q}{(2\pi)^3}\frac{q_i}{8E_{D^{*0}}E_{\bar D^{*0}}E_{D^{0}}}\frac{1}{W-E_{\bar D^{*0}}-E_{D^{*0}}+i\epsilon}\frac{1}{W-E_{\bar D^{*0}}-k^0-E_{D^{0}}+i\epsilon}=k_iA,\\
 &A=\int \frac{d^3q}{(2\pi)^3}\frac{\vec{q}\cdot\vec{k}}{\vec{k}^2}\frac{1}{8E_{D^{*0}}E_{\bar D^{*0}}E_{D^{0}}}\frac{1}{W-E_{\bar D^{*0}}-E_{D^{*0}}+i\epsilon}\frac{1}{W-E_{\bar D^{*0}}-k^0-E_{D^{0}}+i\epsilon}.
\end{align}
Then,
\begin{align}
 (t_T)_i=&k_i\int\frac{2\pi q^2dqd\cos\theta}{(2\pi)^3}\frac{(2+\frac{\vec{q}\cdot\vec{k}}{\vec{k}^2})}{8E_{D^{*0}}E_{\bar D^{*0}}E_{D^{0}}}\frac{1}{W-E_{\bar D^{*0}}-E_{D^{*0}}+i\epsilon}\frac{1}{W-E_{\bar D^{*0}}-k^0-E_{D^{0}}+i\epsilon}\notag\\
 \equiv&k_i \tilde{t}_T. \label{eq:tTtil}
\end{align}
Noticing that the integral is ultraviolet convergent, the integrations of $q$ and $\cos\theta$ can be evaluated numerically in a straightforward way.

Finally, we obtain the $B^-\to K^-\pi^0X(3872)$ decay amplitude given by the triangle diagrams in Figs.~\ref{fig1}a) and \ref{fig1}b) as
\begin{align}
 \mM_{B^-\to K^-\pi^0X}=-\frac{g_X\tilde{g}\mC}{\sqrt{2}}(\vec{\epsilon}_X\cdot\vec{k})\tilde{t}_T,
\end{align}
and the $\pi^0X$ invariant mass distribution is written as
\begin{align}
 \frac{d\Gamma_{B^-\to K^-\pi^0X}}{dM_{\pi^0X}}(M_{\pi^0X},M_X)
 =&\frac{p_{K^-}\tilde{p}_{\pi^0}}{(2\pi)^34M_{B^-}^2}\left(\frac{g_X\tilde{g}\mC}{\sqrt{2}}\right)^2k^2|\tilde{t}_T|^2\label{eq:distnoconv}
\end{align}
with
\begin{align}
 p_{K^-}=&\frac{1}{2M_{B^-}}\lambda^{1/2}(M_{B^-}^2,m_{K^-}^2,M_{\pi^0X}^2),\\
 \tilde{p}_{\pi^0}=k=&\frac{1}{2M_{\pi^0X}}\lambda^{1/2}(M_{\pi^0X}^2,m_{\pi^0}^2,M_{X}^2).
\end{align}

We should note that the $X(3872)$ does not have a precise mass since it has a width, and, hence, it has a mass distribution (spectral function) 
that has to be taken into account for a realistic comparison with experiment (see related work in Ref.~\cite{achapion}).

With a spectral function of the $X(3872)$, $\rho_X(m)$,
\begin{align}
 \rho_X(m)=&\frac{1}{\mN}\left(-\frac{1}{\pi}\right)\im\left[\frac{1}{m^2-{M}_X^2+i{M}_X{\Gamma}_X}\right],\\
 \mN=&\int_{{M}_X-2{\Gamma}_X}^{{M}_X+2{\Gamma}_X} dm 2m \left(-\frac{1}{\pi}\right)\im\left[\frac{1}{m^2-{M}_X^2+i{M}_X{\Gamma}_X}\right], 
\end{align} 
where ${M}_X$ and ${\Gamma}_X$ are the nominal mass and width of $X(3872)$,
the 
$\pi^0X$ 
mass distribution convoluted with the $X(3872)$ spectral function is given by
\begin{align}
 \overline{\frac{d\Gamma_{B^-\to K^-\pi^0X}}{dM_{\pi^0X}}}(M_{\pi^0X})=&\int_{{m}_X-2{\Gamma}_X}^{{m}_X+2{\Gamma}_X} dm 2m \rho_X(m){\frac{d\Gamma_{B^-\to K^-\pi^0X}}{dM_{\pi^0X}}}(M_{\pi^0X},m).\label{eq:dgdmbar}
\end{align}

\section{Results and discussion}
\label{sec:res}

In Fig.~\ref{fig8}, we show the plot of $\overline{{d\Gamma_{B^-\to K^-\pi^0X}}/{dM_{\pi^0X}}}/\Gamma_{B^-}$ as a function of the $\pi^0X(3872)$ invariant mass, $M_{\pi^0X}$, given in Eq.~\eqref{eq:dgdmbar} with $\delta_X=\pm 150$, $\pm 100$, $\pm 50$, and $0~\kev$  and ${\Gamma}_X=100$~keV.
\begin{figure}[t]
 \centering
 \includegraphics[width=12cm]{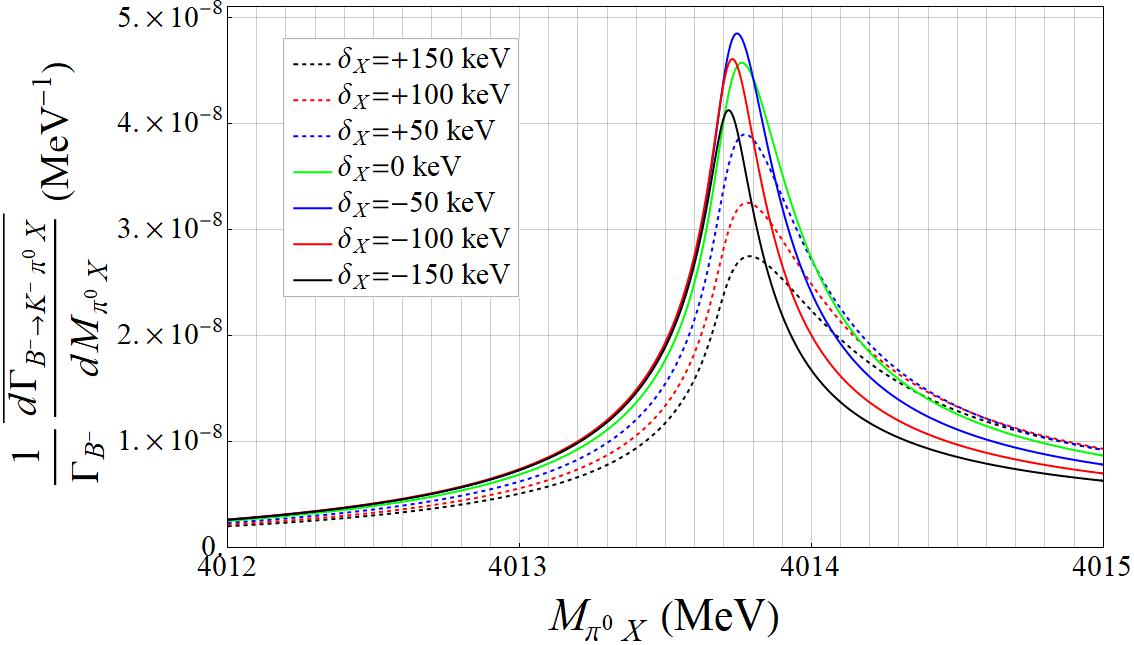}
 \caption{Plot of $\overline{{d\Gamma_{B^-\to K^-\pi^0X}}/{dM_{\pi^0X}}}/\Gamma_{B^-}$ as a function of $M_{\pi^0X}$ with $\Gamma_X=100~\kev$ and $\delta_X=\pm 150$, $\pm 100$, $\pm 50$, and $0~\kev$. 
 }
 \label{fig8} 
\end{figure}
In Fig.~\ref{fig4}, we show the plot of the same results but normalized to the same value at the peak
with the maximum of the $\delta_X=0~\kev$ case. 
\begin{figure}[t]
 \centering
 \includegraphics[width=12cm]{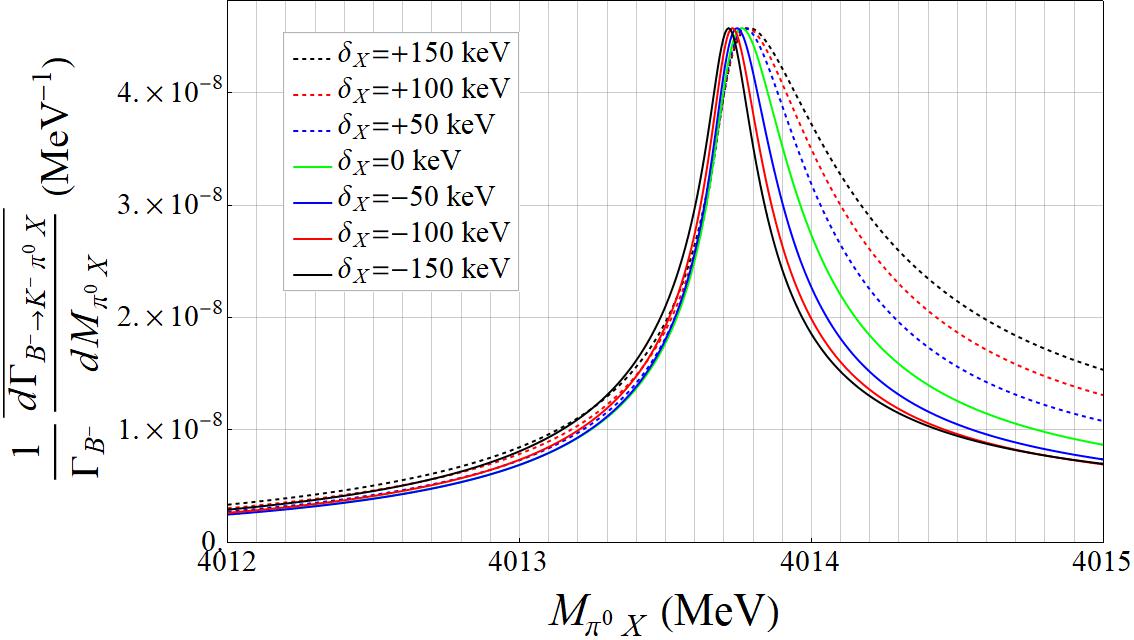}
 \caption{Plot of $\overline{{d\Gamma_{B^-\to K^-\pi^0X}}/{dM_{\pi^0X}}}/\Gamma_{B^-}$ as a function of $M_{\pi^0X}$ with $\Gamma_X=100~\kev$ and $\delta_X=\pm 150$, $\pm 100$, $\pm 50$, and $0~\kev$.
 The lines are normalized with the maximum of that for the $\delta_X=0~\kev$ case.
 }
 \label{fig4} 
\end{figure}
The plot in the smaller range of $M_{\pi^0X}$ $(M_{\pi^0X}\in[4013.5,4014.5]~\mev)$ is given in Fig.~\ref{fig5}.
The lines are normalized with the maximum of the $\delta_X=0~\kev$ case again.
\begin{figure}[t]
 \centering
 \includegraphics[width=13cm]{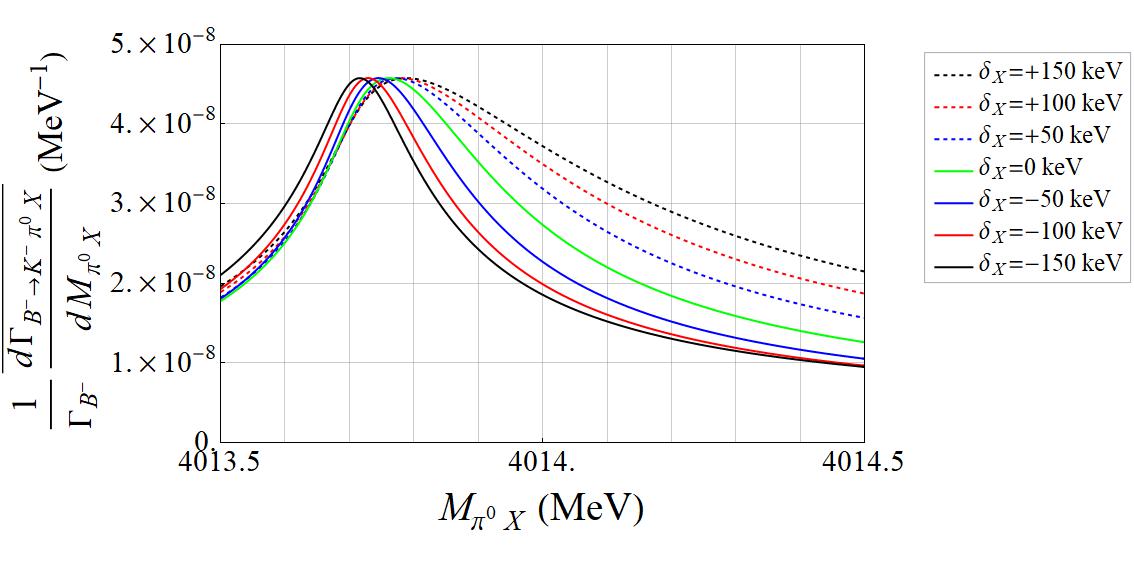}
 \caption{Plot of $\overline{{d\Gamma_{B^-\to K^-\pi^0X}}/{dM_{\pi^0X}}}/\Gamma_{B^-}$ as a function of $M_{\pi^0X}$ with $\Gamma_X=100~\kev$ and $\delta_X=\pm 150$, $\pm 100$, $\pm 50$, and $0~\kev$ in $m_{\pi^0X}\in[4013.5,4014.5]~\mev$.
 The lines are normalized with the maximum of $\delta_X=0~\kev$.
 }
 \label{fig5} 
\end{figure}

In Figs.~\ref{fig9}, \ref{fig6}, and \ref{fig7}, we show the same results as in Figs.~\ref{fig8}, \ref{fig4}, and \ref{fig5}, respectively, but this time with ${\Gamma}_X=0$, i.e., the mass distribution given by Eq.~\eqref{eq:distnoconv}.
\begin{figure}[t]
 \centering
 \includegraphics[width=12cm]{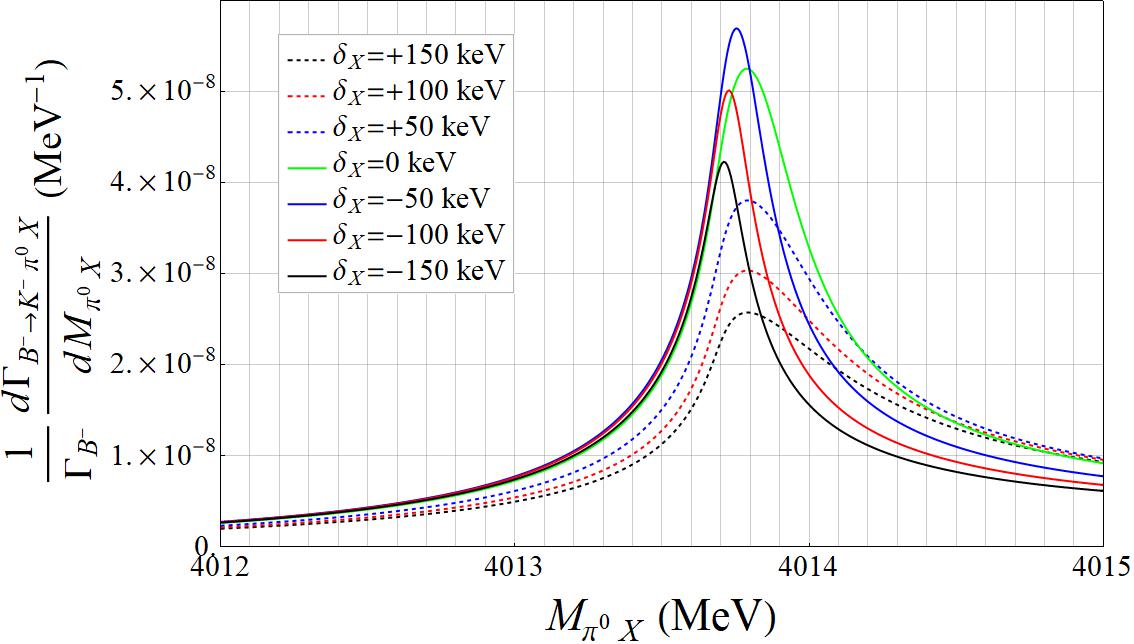}
 \caption{Plot of ${{d\Gamma_{B^-\to K^-\pi^0X}}/{dM_{\pi^0X}}}/\Gamma_{B^-}$ as a function of $M_{\pi^0X}$ with $\delta_X=\pm 150$, $\pm 100$, $\pm 50$, and $0~\kev$.
 The width of $X(3872)$ is not taken into account.
 }
 \label{fig9}
\end{figure}
\begin{figure}[t]
 \centering
 \includegraphics[width=12cm]{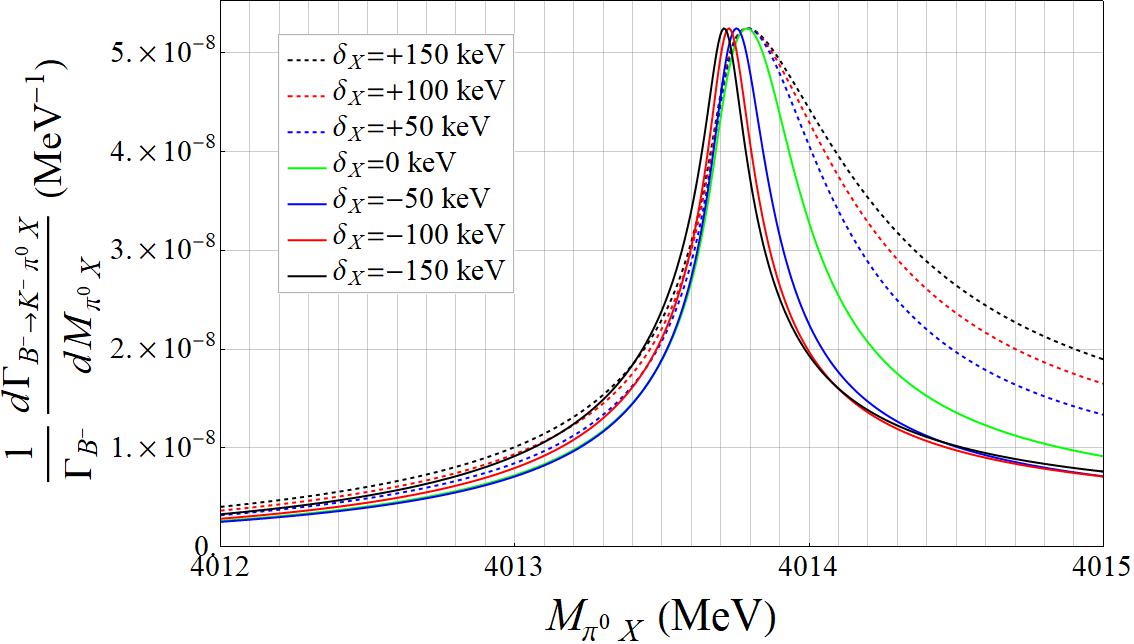}
 \caption{Plot of ${{d\Gamma_{B^-\to K^-\pi^0X}}/{dM_{\pi^0X}}}/\Gamma_{B^-}$ as a function of $M_{\pi^0X}$ with $\delta_X=\pm 150$, $\pm 100$, $\pm 50$, and $0~\kev$.
 The width of $X(3872)$ is not taken into account.
 The lines are normalized with the maximum of $\delta_X=0~\kev$.
 }
 \label{fig6}
\end{figure}
\begin{figure}[t]
 \centering
 \includegraphics[width=13cm]{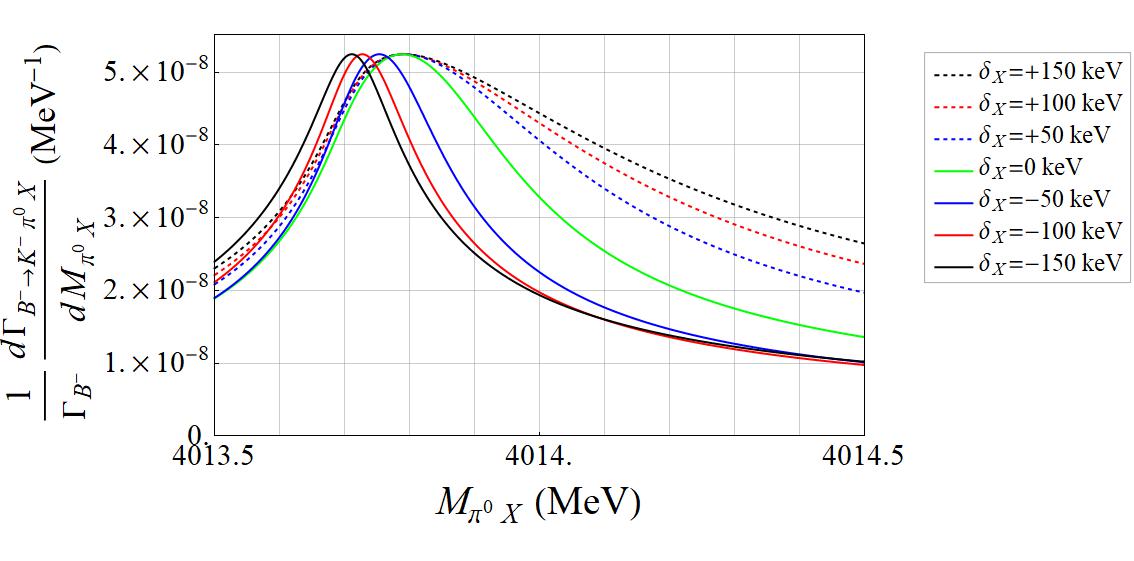}
 \caption{Plot of ${{d\Gamma_{B^-\to K^-\pi^0X}}/{dM_{\pi^0X}}}/\Gamma_{B^-}$ as a function of $M_{\pi^0X}$ with $\delta_X=\pm 150$, $\pm 100$, $\pm 50$, and $0~\kev$ in $M_{\pi^0X}\in[4013.5,4014.5]~\mev$.
 The width of $X(3872)$ is not taken into account.
 The lines are normalized with the maximum of $\delta_X=0~\kev$.}
 \label{fig7} 
\end{figure}
We can see that one effect of considering the width of 100~keV in the $X(3872)$ mass distribution is a reduction of the strength, but the peak position and the widths are very similar.

We should note that not only the peak position but also, more importantly, the asymmetric line shapes of the distributions are rather different for different $M_X$ values.
For instance, by looking at Fig.~\ref{fig5}, the differences in shape between $\delta_X=50$~keV and $-50$~keV are perfectly visible assuming experimental errors of even 20\%.

Integrating the $\pi^0X$ invariant mass distribution in the range of the plot ($M_{\pi^0X}\in[4012,\,4015]~\mev$), we obtain the branching fraction of $B^-\to K^-\pi^0X(3872)$ in this region via a TS to be of $\mathcal{O}(4\times10^{-8})$.

In Fig.~\ref{fig5}, the sensitivity of the peak of the TS to the $X(3872)$ binding energy is clear.
We can see that changes in the binding energy of 100~keV revert into similar changes in the position of the TS peak in $M_{\pi^0X}$, but, even more striking, the shapes are evidently different.
The interesting thing is that one does not need to measure the $X(3872)$ nor the $\pi^0$.
It is sufficient to know that a $\pi^0$ has been produced but its precise measurement is unnecessary.
Technically, it is not even necessary to measure the $\pi^0$, but its simultaneous detection together with the $K^-$ reduces drastically the background.
One measures the $K^-$ energy and determines
\begin{align}
 M_{\pi^0X}^2=(p_B-p_K)^2=M_{B^-}^2+m_{K^-}^2-2M_{B^-}E_{K^-}.
\end{align}
Thus, a precise measurement of the $K^-$ energy for a $B^-$ at rest for the case where the peak appears is all that is needed to determine $M_X$, establishing the correspondence of the peak observed and $M_X$ in Fig.~\ref{fig5}.

We should note that the convolution with the spectral function of the $X(3872)$ considering its width smooths the peaks and changes the shapes.
We have seen that if we take $\Gamma_X=1.2$~MeV, which is the current upper limit, the dependency of the peak position of the TS with $M_X$ would be softened.
However, there are good reasons to think that the width of the $X(3872)$ would not be larger than $100$~keV,
which makes our predictions realistic: the $D^0\bar D^0\pi^0$ mode consists of $\gtrsim40\%$ branching fraction of the $X(3872)$ decays~\cite{Gokhroo:2006bt,pdg}, and its partial decay width is expected to be about $40$~keV~\cite{mehen,guo1,guo2}.
The rates obtained are small, but $B^-$ branching ratios of the order of $10^{-7}$ are already recorded~\cite{pdg}.
One should also note that the $B^+\to K\pi X(3872)$ has been measured (not the specific TS peaks) with ratios $(1.06\pm 0.31)\tento{-5}$ for $\br(B^+\to K^0\pi^+X(3872))\times \br(X(3872)\to J/\psi \pi^+\pi^-)$ \cite{belle} and $(7.9\pm 1.3\pm 0.4)\tento{-6}$ for $\br(B^0\to K^+X\pi^-)\times \br(X(3872)\to J/\psi\pi^+\pi^-)$~\cite{belle}.
Using the branching ratio for $X(3872)\to J/\psi \pi^+\pi^-$ of the order of 4\% in Refs.~\cite{changzheng,babar}, this gives $\br(B^+\to K^0\pi^+ X(3872))$ of the order of $2\tento{-4}$. 
In addition, in the low $\pi X$ region, the TS contribution is more important for the $B^-\to K^-\pi^0X(3872)$ than for the $B^+\to K^0\pi^+X(3872)$~\cite{brapion}.
Future updates of the present facilities should make the rates of $\mathcal{O}(4\times10^{-8})$ attainable.

We have mentioned above the sensitivity of the shape above the peak to the $X$ binding energy.
In order to make this point more clear, we define the following asymmetry:
\begin{equation}
    \frac{N_>}{N_<} \equiv \frac{ \int_{M^\text{max}}^{M^\text{max} +\delta}d M_{\pi^0 X} \left(\overline{d\Gamma_{B^-\to K^-\pi^0X}/{dM_{\pi^0X}}} \right) }{ \int^{M^\text{max}}_{M^\text{max} -\delta}d M_{\pi^0 X} \left(\overline{d\Gamma_{B^-\to K^-\pi^0X}/{dM_{\pi^0X}}} \right) }, 
    \label{eq:asy}
\end{equation}
with $M^\text{max}$ being the $\pi^0X$ invariant mass where the $\pi^0X$ line shape takes its maximal value and $\delta$ a small range, say, 2~MeV. The dependence of this asymmetry on the $X$ binding energy $\delta_X$ is shown in Fig.~\ref{fig:NgrovNlt}. 
We find that this magnitude is very sensitive to changes of $\delta_X$.
Since one has more statistics in an integrated distribution, this magnitude could turn out to be the ideal one to determine the $X$ mass with precision, and
the $X(3872)$ binding energy may be extracted by measuring this asymmetry.
\begin{figure}[t]
    \centering
    \includegraphics[width=10cm]{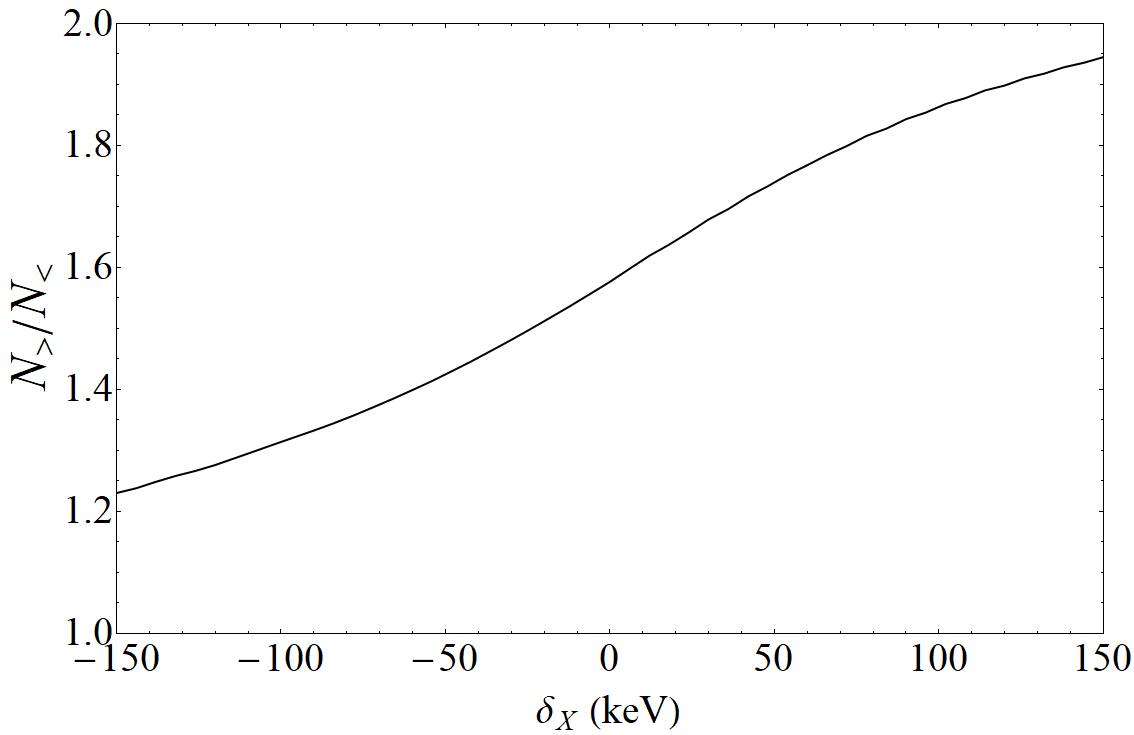}
    \caption{Dependence of the asymmetry $N_>/N_<$ defined in Eq.~\eqref{eq:asy} on the $X(3872)$ binding energy $\delta_X$.}
    \label{fig:NgrovNlt}
\end{figure}

\section{Conclusions}
\label{sec:conc}

We have made a study of the $B^- \to K^- \pi^0 X(3872)$ decay considering a triangle mechanism with a primary $B^- \to K^- D^{*0}\bar D^{*0}$ decay followed by $D^{*0}\to\pi^0 D^0$ and the fusion of $D^0\bar D^{*0}$ to give the $X(3872)$.
The triangle diagram of this mechanism develops a triangle singularity in a very narrow window of $X(3872)$ masses, 
between 3871.68~MeV and 3971.93~MeV, producing a peak in the $\pi^0 X$ invariant mass around 4013.75~MeV.
The results of our investigation point out the high sensitivity of the peak position and, in particular, the line shape of the $\pi^0 X$ mass distribution of this process to the $X(3872)$ binding energy.
The shape predicted in this work is very useful to determine the $X(3872)$ mass, 
and we show that even with 20\% uncertainties in the measured points of the mass distribution, one could discriminate 100~keV in the $X(3872)$ mass relative to the $D^0\bar D^{*0}$ threshold.  
To stress this property, we introduced a new magnitude, meaning the asymmetry of the distribution, and found that it is very sensitive to the $X(3872)$ mass.
This benefits with larger statistics since it involves integrated mass distributions.
The novelty in the reaction proposed is that it establishes a one-to-one correspondence between the $X(3872)$ binding energy and the asymmetry in the line shape of the $\pi^0 X$ mass distribution. 
Technically, one does not need to measure the $\pi^0$. 
The $\pi^0 X$ invariant mass would be determined from the energy of the emitted kaon, which can be measured with high precision. 
Yet, the detection of the pion, albeit without the need for precision, would serve to reduce the background. 
  
The obtained branching ratios are in the limit of the smallest present rates reported by the PDG and should be accessible in new rounds of measurements or future updates of the present facilities. 
With a more precise measurement of the $X(3872)$ binding energy and its width, both of which are intimately connected to the structure of the $X(3872)$, new insights into the nature of the $X(3872)$, which is the object of much debate, are foreseen.

\medskip

\section*{Acknowledgements}

F.-K.G. is grateful to the hospitality of the Helmholtz Institut f\"ur Strahlen- und Kernphysik where part of the work was done.
This work is supported in part by the National Natural Science Foundation of China (NSFC) and  the Deutsche Forschungsgemeinschaft (DFG) through the funds provided to the Sino-German Collaborative Research Center  CRC110 ``Symmetries and the Emergence of Structure in QCD"  (NSFC Grant No. 11621131001), by the NSFC under Grants No. 11835015, No.~11947302 and No. 11961141012, by the Chinese Academy of Sciences (CAS) under Grants No. QYZDB-SSW-SYS013 and No. XDPB09, by the CAS Center for Excellence in Particle Physics (CCEPP), by the Spanish Ministerio de Economia y Competitividad and European FEDER funds under Contracts No.~FIS2017-84038-C2-1-P B and No.~FIS2017-84038-C2-2-P B, by the Generalitat Valenciana in the program Prometeo II-2014/068, and by the project Severo Ochoa of Instituto de F\'isica Corpuscular (IFIC), SEV-2014-0398.
This project has received funding from the European Union's Horizon 2020 research and innovation programme under grant agreement No.~824093 for the STRONG-2020 project.
S.S. is also supported by the 2019 International Postdoctoral Exchange Program and by the CAS President's International Fellowship Initiative (PIFI) under Grant No. 2019PM0108.

\medskip

\bibliography{biblio}

\end{document}